\documentclass[letterpaper,twocolumn,10pt]{article}
\usepackage{usenix2019_v3} %
\usepackage{adjustbox}
\usepackage{amssymb}
\usepackage{amsmath}
\usepackage{multicol}
\usepackage{supertabular}
\usepackage{tikz,graphicx,balance,multirow,xspace,subcaption,longtable}
\usepackage{makecell,amsmath,cleveref,rotating}
\usepackage[T1]{fontenc}
\usepackage{enumitem}
\usepackage[draft,inline,nomargin,index]{fixme}
\usepackage{booktabs}
\usepackage{lastpage}
\usepackage{titling}
\usepackage{xcolor,colortbl}
\usepackage{diagbox}
\usepackage{enumitem}
\usepackage{wasysym}

\setitemize{noitemsep,topsep=0pt,parsep=0pt,partopsep=0pt}

\usepackage[font={small}, textfont=md, skip=.2pt]{caption} %
\PassOptionsToPackage{hyphens}{url}
\usepackage{hyperref}
\newcolumntype{P}[1]{>{\arraybackslash}p{#1}}
\newcolumntype{X}[1]{>{\centering\arraybackslash}p{#1}}

\expandafter\def\expandafter\UrlBreaks\expandafter{\UrlBreaks%
  \do\a\do\b\do\c\do\d\do\e\do\f\do\g\do\h\do\i\do\j%
  \do\k\do\l\do\m\do\n\do\o\do\p\do\q\do\r\do\s\do\t%
  \do\u\do\v\do\w\do\x\do\y\do\z\do\A\do\B\do\C\do\D%
  \do\E\do\F\do\G\do\H\do\I\do\J\do\K\do\L\do\M\do\N%
  \do\O\do\P\do\Q\do\R\do\S\do\T\do\U\do\V\do\W\do\X%
  \do\Y\do\Z}

\crefformat{section}{\S#2#1#3}
\crefformat{subsection}{\S#2#1#3}
\crefformat{subsubsection}{\S#2#1#3}

\fxsetup{theme=colorsig,mode=multiuser,inlineface=\itshape,envface=\itshape}
\FXRegisterAuthor{ht}{aht}{Hammas}
\FXRegisterAuthor{rn}{arn}{Rishab}
\FXRegisterAuthor{rs}{ars}{Rachee}
\FXRegisterAuthor{pp}{app}{Paul}

\hypersetup{%
  pdftitle={},
  pdfauthor={},
  pdfkeywords={},
  bookmarksnumbered,
  pdfstartview={FitH},
  colorlinks,
  citecolor=black,
  filecolor=black,
  linkcolor=black,
  urlcolor=blue,
  breaklinks=true,
  hypertexnames=false
}

\newcommand\clearrow{\global\let\rowmac\relax}

\newcommand{\para}[1]{{\vspace{.05in} \bf \noindent #1 }}

\newcommand{\red}[1]{\textcolor{red}{\bf #1}}

\newcommand{\etal}{et al.\xspace}
\newcommand{\ie}{i.e.,\ }

\definecolor{red00}{rgb}{1,0.9,0.9}
\definecolor{red0}{rgb}{1,0.8,0.8}
\definecolor{red1}{rgb}{1,0.7,0.7}
\definecolor{red2}{rgb}{1,0.6,0.6}
\definecolor{red3}{rgb}{1,0.5,0.5}
\definecolor{red4}{rgb}{1,0.4,0.4}
\definecolor{red5}{rgb}{1,0.3,0.3}
\definecolor{green0}{rgb}{0.8,1,0.8}
\definecolor{green1}{rgb}{0.7,1,0.7}
\definecolor{green2}{rgb}{0.6,1,0.6}
\definecolor{green3}{rgb}{0.5,1,0.5}
\definecolor{green4}{rgb}{0.4,1,0.4}
\definecolor{green5}{rgb}{0.3,1,0.3}

\newcommand{\FigProbeSend}{
    \begin{figure}[t]
     \centering
     \includegraphics[width=0.95\linewidth]{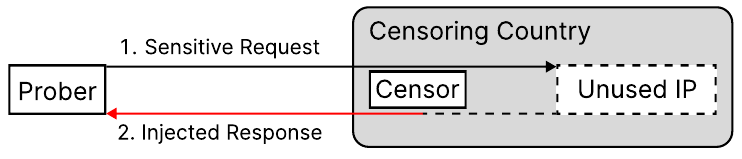}
     \caption{\textbf{Sending Probes}\,---\,%
            For each protocol we test, we send probes to addresses from which we do not expect responses. When a passive censor monitoring traffic along the
            path to a target address is triggered by a blocklisted domain it
            injects a packet where no response would exist otherwise.}
    \label{fig:probeSend}
    \end{figure}
}

\newcommand{\FigAllocSize}{
    \begin{figure*}[t]
     \centering
     \includegraphics[width=0.95\linewidth]{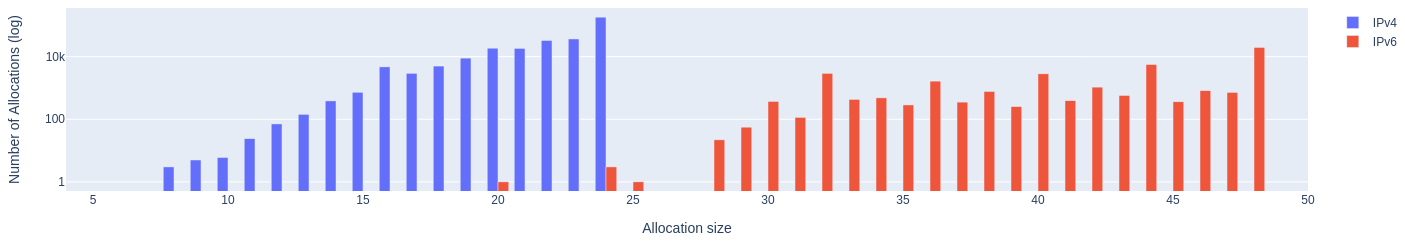}
     \caption{\textbf{Target allocation sizes}\,---\,%
            We draw our target addresses at random from BGP announced
            allocations. This graph shows the allocation sizes from which we
            select for both IPv4 and IPv6.}
    \label{fig:allocsizes}
    \end{figure*}
}

\newcommand{\PrevalenceGeneral}{
  \begin{table*}
    \centering
    \begin{tabular}{|c|cc|cc|cc|cc|cc|cc|}
    \hline
    \diagbox{\textbf{Country}}{\textbf{Protocol}} & \multicolumn{2}{c|}{\begin{tabular}[c]{@{}c@{}}\textbf{~DNS A~}\\\textit{\small{IPv4~~~IPv6}}\end{tabular}} & \multicolumn{2}{c|}{\begin{tabular}[c]{@{}c@{}}\textbf{~ ~ DNS~ ~~}\\\textbf{AAAA}\end{tabular}} & \multicolumn{2}{c|}{\begin{tabular}[c]{@{}c@{}}\textbf{HTTP }\\\textbf{(stateful)}\end{tabular}} & \multicolumn{2}{c|}{\textbf{~ ~HTTP~ ~}} & \multicolumn{2}{c|}{\begin{tabular}[c]{@{}c@{}}\textbf{TLS }\\\textbf{(stateful)}\end{tabular}} & \multicolumn{2}{c|}{\textbf{TLS}}  \\
    \hline
    \textbf{China}                                & 87 & 99                                                                                     & 87 & 99                                                                                          & 69 & 99                                                                                          &  & 61                                  & 76 & 99                                                                                         &  & 50                            \\
    \textbf{Iran}                                 & 71 & 32                                                                                     & 71 & 32                                                                                          & 69 & 32                                                                                          & 39 & 31                                  &  & 32                                                                                         &  & 26                            \\
    \textbf{Uzbekistan}                           &    &                                                                                        &    &                                                                                             & 90 & 83                                                                                          &   &                                   & 91 & 83                                                                                         &   &                             \\
    \textbf{Oman}                                 &    &                                                                                        &    &                                                                                             & 83 & 70                                                                                          & 83 & 90                                  & 81 & 70                                                                                         &   &                               \\
    \textbf{Morocco}                              &    &                                                                                        &    &                                                                                             & 63 &                                                                                             &  &                                     &  &                                                                                            &   &                               \\
    \textbf{Bangladesh}                           &    &                                                                                        &    &                                                                                             & 59 & 71                                                                                          & 60 & 71                                  &    &                                                                                            &    &                               \\
    \textbf{Tanzania}                             &    &                                                                                        &    &                                                                                             & 41 &                                                                                             & 35 &                                     &    &                                                                                            &    &                               \\
    \textbf{Kuwait}                               &    &                                                                                        &    &                                                                                             & 40 &                                                                                            &    &                                     &    &                                                                                            &    &                               \\
    \textbf{Libya}                                &    &                                                                                        &    &                                                                                             & 27 & 100                                                                                         & 27 & 100                                 & 27 & 100                                                                                        &    &                               \\
    \textbf{Pakistan}                             &    &                                                                                        &    &                                                                                             & 25 & 29                                                                                          &    &                                     & 25 & 27                                                                                         &    &                               \\
    \textbf{Lebanon}                              &    &                                                                                        &    &                                                                                             & 20 &                                                                                             &    &                                     & 20 &                                                                                            &    &                               \\
    \textbf{Turkey}                               &    &                                                                                        &    &                                                                                             & 20 &                                                                                             &    &                                     &  &                                                                                            &    &                               \\
    \hline
    \end{tabular}
    \caption{\textbf{Global view of bidirectional censorship by protocol}\,---\,
        For each country and protocol,
        we represent the pair of percentage of vantage points in that country (IPv4 and IPv6 respectively) that returned a censorship result.
        We omit numbers if they are below a threshold of 5\% for both IPv4 and IPv6, indicating that our bidirectional censorship measurement technique did not observe that protocol blocked in the specified country.
    }
    \label{fig:prevalencegeneral}
    \end{table*}
}

\newcommand{\TabQuack}{\begin{table}
    \centering
    \begin{tabular}{l|ccc}
       & Shared & & \\ 
       & Allocations & Quack & Our work  \\
    \hline
    CN & 373                & 98\%  & 59\%       \\
    RU & 139                & 26\%  & 3\%        \\
    PK & 12                 & 92\%  & 0\%        \\
    TR & 10                 & 40\%  & 0\%        \\
    LB & 5                  & 0\%   & 0\%        \\
    IR & 1                  & 100\% & 0\%        \\
    KW & 1                  & 0\%   & 0\%

    \end{tabular}
    \caption{\textbf{\textbf{Quack Comparison}\,---\,} We compared our bidirectional
  censorship measurements to those obtained by
  Quack~\cite{vandersloot2018quack}. We looked at allocations that had both a
  Quack echo server and a measurement from our study, and report on the fraction
  of those vantage points that Quack or we would label as censored.
  We observe that in many countries, we do
  not observe bidirectional censorship even in subnets that Quack sees
  significant blocking. Since Quack can detect censorship that occurs only for
  connections originating from within the country, this data supports our
  hypothesis that these countries do not censor bidirectionally.
  }
    \label{tab:quack}
    \end{table}
}

\begin{document}

\pagestyle{plain} %

\renewcommand{\sectionautorefname}{\S}
\renewcommand{\subsectionautorefname}{\S}
\renewcommand{\subsubsectionautorefname}{\S}
\date{}

\setlength{\droptitle}{-5em}   %
\posttitle{\par\end{center}}   %
\title{{\bf ProtoScan}\\\Large{Measuring censorship in IPv6}}
\author{
{\rm Jack Wampler }\\
University of Colorado Boulder
\and
{\rm Hammas Bin Tanveer}\\
University of Iowa
\and
{\rm Rishab Nithyanand}\\
University of Iowa
\and
{\rm Eric Wustrow}\\
University of Colorado Boulder
}

\maketitle

\begin{abstract}

Internet censorship continues to impact billions of people worldwide,
and measurement of it remains an important focus of research.
However, most Internet censorship measurements have focused solely on the IPv4
Internet infrastructure. Yet, more clients and servers are available over IPv6:
According to Google, over a third of their users now have native IPv6 access.

Given the slow-but-steady rate of IPv6 adoption, it is important to understand
its impact on censorship. In this paper, we measure and analyze how censorship
differs over IPv6 compared to the well-studied IPv4 censorship systems in use
today.

We perform a comprehensive global study of censorship across an array of
commonly censored protocols, including HTTP, DNS, and TLS, on both IPv4
and IPv6, and compare the results.
We find that there are several differences in how countries censor IPv6 traffic,
both in terms of IPv6 resources, and in where and what blocklists or technology
are deployed on IPv6 networks. Many of these differences are
not all-or-nothing: we find that most censors have some capacity to block
in IPv6, but are less comprehensive or less reliable compared to their IPv4
censorship systems.

Our results suggest that IPv6 offers new areas for censorship circumvention
researchers to explore, providing potentially new ways to evade censors. As more
users gain access to IPv6 addresses and networks, there will be a need for tools
that take advantage of IPv6 techniques and infrastructure to bypass censorship.

 \end{abstract}

\section{Introduction}\label{sec:intro}

Internet censorship is a global problem that affects over half the world's
population. Censors rely on sophisticated network middleboxes to inspect and
block traffic, employing IP-based blocking and packet injection to prevent
access to censored content and resources. A common technique used by censors involves
inspecting network traffic
passively, and injecting (spoofed) responses to DNS, TLS, HTTP, or other protocol
requests for censored
content~\cite{lowe2007great,vandersloot2018quack,xu2011internet,aryan2013internet,chai2019importance,elmenhorst2021web}.

Prior work has extensively studied this type of censorship
globally~\cite{niaki2020iclab,sundara2020censored,filasto2012ooni,razaghpanah2016exploring,kuhrer2015going,dagon2008corrupted,pearce2017global,scott2016satellite}
and for individual
countries~\cite{Anonymous2020:TripletCensors,USESEC21:GFWatch,aryan2013internet,ramesh2020decentralized,yadav2018light,gebhart2017internet,nabi2013anatomy}.
These studies generally perform active measurements that test large
sets of domains in requests into censored countries, and identify forged
censorship responses from legitimate ones.
Unfortunately, this prior work has focused exclusively on the IPv4 Internet, in
part because scanning the IPv6 Internet for servers is
difficult~\cite{murdock2017target}, owing to its impossible-to-enumerate 128-bit
address space.

An IPv4-only view of censorship is problematic, because
IPv6 is becoming more widely deployed and used worldwide: over 35\% of current Internet
traffic is being served over native IPv6 connections~\cite{Google-IPv6} (and exceeds
50\% in some countries known to censor such as India~\cite{akamai-ipv6}).
However, it
is unclear if the same censorship mechanisms we know about in IPv4 traffic
also apply to the growing IPv6 Internet.
There is also reason to believe it could be
different, as prior work studying IPv6 in non-censorship contexts has shown IPv6 has
fundamentally different network performance~\cite{Dhamdhere-IMC2012}, security
policies~\cite{Czyz-NDSS2016}, and topologies~\cite{Czyz-SIGCOMM2014}
compared to the traditional IPv4 Internet.

In this paper, we perform the first (to our knowledge) comprehensive global measurement of
censorship on the IPv6 Internet, and compare it to IPv4 censorship.
To study censorship globally on both IPv4 and IPv6 networks, we focus on
detecting \emph{bidirectional censorship}, which can be easily observed from a
vantage point outside the country. In this form of censorship, a censor
passively watches network traffic for censored requests, such as a DNS query
for a blocked domain. When the censor sees such a request, they inject a response (such as a DNS
response with an incorrect answer), spoofing the source of the injected
response,
as shown in Figure~\ref{fig:probeSend}.
This type of censorship can be induced and detected from a single
vantage point outside any censoring
country~\cite{vandersloot2018quack,collateral-dns,pearce2017global,scott2016satellite}.
While this technique does not capture other types of censorship (e.g. IP
blocking), it provides one view
of censorship that we can easily apply globally and across network types.

In particular, we randomly sample IP addresses from both IPv4 and IPv6
allocations, with the goal of finding routed-but-unused addresses in every
country. This removes the need to scan for active servers, which is difficult in
IPv6. By looking for unused addresses that don't respond to our control probes,
we can simply sample addresses that we know route into a country of interest.
For each IP, we send requests for several protocols (DNS, TLS, and HTTP)
containing potentially censored domains and observe any injected response. If a
country has a largely uniform response to censored probes, we can label it as
censoring for that protocol and network type (IPv4 or IPv6).

\medskip
Our results show that
some censors such as Tanzania and Turkey only support censorship of
their IPv4 Internet,
while others including China and Iran support both IPv4 and IPv6 censorship.
However, even censors that censor both IPv4 and IPv6 may have subtle differences
between the two: the censorship may apply to fewer networks, may miss certain
kinds of tunneling, apply to different
protocols, or to different domains or resources. These differences may
potentially be useful to circumvention researchers, providing information about
the censorship infrastructure and ways to get around it.

\section{Datasets \& Methodology}\label{sec:methodology}

In this section we outline our technique for measuring bidirectional censorship
globally. While this technique can miss several kinds of censorship (e.g.
censorship limited to a country, or IP-based blocks), it allows us to
efficiently measure censorship around the world from a single vantage point.
Bidirectional censorship occurs when a country's firewall is agnostic to the
direction that packets cross the border, and injects responses even if the
offending request or connection originates outside the country. This allows us
to send a censored packet into the country from our vantage point in
North America, and receive back injected responses.

\FigProbeSend

\subsection{Selecting target domains}
\label{sec:methodology:domains}
We begin by selecting a list of domains that are likely to trigger injected
responses from censoring countries around the world.
We use The Citizen Lab's~\cite{TheCitiz6:online} domain test
list~\cite{testlist}, which includes domains that are known to be blocked by
many censors. We use the global test list (composed of 1397 unique domains) for
our measurements, rather than country-specific lists, to keep our experiments
consistent across countries.

We supplement this list with 10 control domains that to our knowledge are not
blocked by any censors. Our control domains are a combination of domains we set up
specifically for these measurements, and domains that have not been registered
(produce an NXDOMAIN).
If an IP responds to requests (e.g. DNS, HTTP,
TLS, etc) containing our control domains, we assume that IP is a legitimate host
or non-censoring firewall (e.g. paywall or corporate firewall), and remove the
IP from our study. Thus, we use control domains to locate \emph{unused} IPs that
don't respond to our control queries, and test for in-network injections to
these IPs.

\subsection{Selecting IP Addresses}
\label{subsec:selecting-ips}

\FigAllocSize

Our goal is to identify destinations that we can send probes to that will route
past a particular country's censorship infrastructure, but not reach a
responsive host. Ideally, our probe either triggers censorship (if it is on path) and
receives an injection, or the probe is dropped by a router or host in the censoring
country. To achieve this, we select IP addresses to scan that are routed, in a
given country, but ultimately
non-responsive to our control domain probes.

We begin by collecting a list of all IP allocations announced by the 5 regional
registries~\cite{herrbisc56:online} containing 155k IPv4 and
63k IPv6 allocations from a total of over 200 countries. We use
MaxMind~\cite{IPGeoloc87:online} to assign the Autonomous System Numbers (ASNs)
and organization names to each of these allocations, as well as geolocate each
IP address to a specific country. We then filter out
organizations that do not have at least one IPv4 and one IPv6 allocation
resulting in 71k IPv4 and 20k IPv6 allocations.

It is uncommon that the entire IP allocation provided by RIRs is announced as
is, often being split and announced in several smaller allocations. Portions of
the allocated IP block sometimes remain unannounced by the organization.
Choosing IP addresses from the unannounced region of an IP block might reduce
the rate at which probes are truly routed into the country in question
potentially resulting in a false negative, under-representing the prevalence of
bidirectional censorship in a country. To increase the confidence that all
chosen IP addresses are routable and lower the chance of this type of false
negative result we chose IP addresses only from the announced prefixes of each
allocation. We used the \textit{University of Oregon Route Views Project}
\cite{RouteVie20:online} to get all the announced prefixes for each of the
allocated IP block, as of Oct~1.~2022. This resulted
in a total of 295,385 IPv4 and 39,572 IPv6 prefixes representing 186 countries.

For each announced prefix, we select a set of $N=10$ addresses at random. We
arrived at this number after a couple of considerations. First, scanning the
entirety of the addresses space for 1400 domains is infeasible for IPv4 and
impossible for IPv6. Second, our aim is to test for on path censorship on the
penultimate hop and not necessarily reach the end hosts themselves. This
technique allows us to test for bidirectional censorship while maintaining the
breadth of our measurements. Choosing 10 addresses from each of the announced
prefix in our allocation dataset results in over 3.3 million IP addresses.

\subsection{Identifying Bidirectional Censorship}
\label{sec:methodology:censorship}

\textbf{\Cref{fig:probeSend}} shows a high-level overview of measuring
bidirectional censorship from a single vantage point outside the censoring
country. For each domain in our test list, we send a probe to each IP address,
for several different protocols, and observe if we receive an (injected)
response. If we receive a response for non-control domains (and no response for
the control domains), we mark the IP address as likely censored. For instance,
sending a DNS request for \texttt{youtube.com} to an IP address in China usually
results in receiving an injected DNS response from China's Great Firewall, while
sending a query for an uncensored domain we would expect no response.

For each experiment, we send a request for a given domain to each IP in our
3.5~million selected IPs before moving to the next domain. This avoids
overwhelming any individual IP address, as each IP receives a probe
approximately every 9~seconds during our scans.

\subsubsection{Protocols}
We scan for censorship in several protocols, including DNS, HTTP, and TLS.

\paragraph{DNS} For each domain from our list, we craft a DNS query for both A
and AAAA records. We note that each of these queries can be sent to an IPv4 or
IPv6 address, allowing us to observe if the censor can process IPv6 packets or
handles AAAA records properly.

\paragraph{HTTP} Plaintext HTTP is often censored if the Host header contains 
a censored domain. We craft a simple GET request with the domain in the Host
header in order to trigger censorship. %
Since HTTP is sent over TCP, censors may track connection
state, expecting to see at least the client's side of a TCP handshake before a
request in order for the censor to inject a response~\cite{bock2021your}. For
this reason, we send two kinds of HTTP probes. First, we send a plain HTTP
request in a single TCP packet with arbitrary sequence and acknowledgement
numbers. This will trigger \emph{stateless} censors that are only watching for
the presence of HTTP requests, regardless of surrounding connection state. In a
second experiment, we send packets that correspond to a client's side of a TCP
connection, namely the \texttt{SYN}, \texttt{ACK}, and finally \texttt{PSH+ACK}
request packets with appropriate sequence numbers. This experiment will trigger
injections from \emph{stateful} censors that expect to see evidence of a
connection before injecting.

\paragraph{TLS} Censors frequently block TLS connections based on the presence
of censored domains in the Client Hello's Server Name Indication (SNI)
extension, which indicates the domain the client is requesting in plaintext. We
craft a TLS Client Hello resembling that sent by very few other tls
implementations - we note that no widespread blocklist (or allowlist) of tls
fingerprints had been applied by censors to general TLS traffic. Similarly to
HTTP, we send both a single ``stateless'' TLS packet, and a separate
``stateful'' \texttt{SYN} / \texttt{ACK} / \texttt{PSH+ACK} sequence to trigger
censors that don't or do track TCP state respectively.

\if{0}
\paragraph{QUIC} QUIC is an encrypted UDP protocol, and like TLS, its first
packet is a Client Hello that conveys the domain in an SNI extension. Unlike
TLS, this first packet is encrypted under a secret derived from a
specification-defined value and the connection ID sent in plaintext. This allows
a network device (e.g. censor) to decrypt the QUIC Client Hello, but requires an
understanding of the QUIC specification. To test if censors are censoring QUIC
or specific domains (and decrypting Client Hello messages), we craft a standard
QUIC Client Hello message with each domain.
\fi

\subsubsection{Controlling for responsive targets and residual censorship}
We exclude IP addresses from our scans that respond to any of our control
domains, since this indicates either a host or firewall that is likely blocking
\emph{all} requests. Typically, this is due to a host being active there,
potentially sending a \texttt{RST} packet in response to our TCP packets. Since
our goal is to measure in-network bidirectional censorship, we exclude such
``live'' hosts from our measurements.

A second related issue we consider is \emph{residual
censorship}~\cite{bock2021your}, where a censor will block a censored request,
and subsequently block future connections from the same client to the same
destination for a short time after, even if those future connections or requests
contained uncensored domains. If we interpret our results naively, residual
censorship could skew our results, making a domain appear to be censored when in
reality only a domain probed shortly before it actually was.

Thus, a limitation of our scanning methodology is that we cannot identify which
domains are blocked by a censor that employs residual censorship. Instead, we
only attempt to identify which IP addresses likely experience censorship, and
then at a country or AS level, what fraction of IPs experience censorship.
In future work, we plan to scan at a slow enough rate or to different IPs in the
same subnet to avoid the residual censorship issue.

\subsubsection{Tagging}

Similar to the architecture of the ZMap~\cite{Durumeric13zmap} scanning tool,
our probing architecture uses many threads to craft and send packets and one
independent thread to listen and ingests responses. One consequence of this
architecture is that we must maintain a limited amount of state internally for
each connection we create. While injected responses to DNS probes may include
the host name (in the response Resource Record) other protocols are not
guaranteed to do so. For example, the TCP RST packets injected by the GFW in
response to a TLS probe with a censored SNI will not indicate what domain the
original probe included. Similar challenges exist for HTTP probes.

To solve this problem we ``tag'' outgoing packets in a way that injected
responses will echo this tag, and allow us to identify the details of the probe
they correlate to, as well as check the
validity of the response without tracking the full connection state from start
to finish.

We start by creating a 1-to-1 mapping from domain to a random number in the
range 1000-65535. We use this number as the source port for the outgoing packet
meaning that we can use the destination port of any response packet to lookup
the domain sent in the original probe. In order to ensure that response TCP
packets are associated with our measurement and not just sent randomly we set
the acknowledgement number of the outgoing probe to be the CRC32 of the source
port (from our domain mapping) and the target address. This allows our ingest
thread to quickly validate responses by checking:
\begin{gather*}
CRC32(PORT_{dst},ADDR_{src}) \stackrel{?}{=} SEQ - Len
\end{gather*}

\if{0}
For Quic responses that either return garbage or change the connection ID for
the server initial packet we need to have access to the 8 byte connection ID
sent in our probe. To make sure this is always available we set the source port
for the outgoing packets from our 1-to-1 domain map and then set the connection
ID to be the CRC64-ECMA of the source port and the target address for the
outgoing packet. That way response packets can statelessly derive the original
connection ID by computing the following for incoming packets.
\begin{gather*}
Conn\_ID = CRC64_{ECMA}(PORT_{dst},ADDR_{src})
\end{gather*}
\fi

\subsection{Ethics}\label{sec:methodology:ethics}

Our experimental design has incorporated ethical considerations into the
decision-making process at multiple stages. Censorship measurement has inherent
risks and trade-offs: better understanding of censorship can help support and
inform users, but specific measurements may carry risk to participants or
network users. Measurement of bidirectional censorship typically allows
researchers to limit the number of third parties implicated in experiments as
the censorship response can be triggered by either ingress or egress traffic
removing the need for cooperation by individual hosts or hosting services within
a censoring region. Vantage points are instead hosted in regions that do not
censor connections and allow for researchers to freely measure the internet.

The vantage that was used for data collection is connected to the internet with
a 1 Gbps interface that scanned using the default rates for our custom protocol
scanning tool (line rate). However, the structure of the scan was established
such that individual addresses and domains would be accessed in round robin
order --- \ie when sending probes every target address would receive a first
request before any target would receive the subsequent request.

We encourage readers to consult The Menlo Report~\cite{menlo}, its companion
guide~\cite{menlo-companion}, and the censorship specific ethical measurement
guidelines outlined by Jones \etal \cite{jones2015ethical} for further
discussion of ethical design for internet measurement.

\section{Results}
\PrevalenceGeneral
\label{sec:prevalence}

In this section, we provide results from our Internet scans, broken down by
country and protocol. For each country and protocol, we label the country as
censoring that protocol (bidirectionally) if more than 20\% of the IPs in that
country returned an injected response for any censored domain queries (and not
for our control domains). For instance, in China, $87.2\%$ of IPv4 addresses we
scanned (that didn't respond to our control queries) provided an injected
response for our DNS A record queries for at least one of the 1400 domains we
tested, confirming that China has a robust and widespread system in place to
censor DNS.

\Cref{fig:prevalencegeneral} shows a breakdown of the countries that we observed
censorship in for the protocols we tested.

\subsection{Prevalence of censorship by protocol}
\label{sec:prevalence:proto}

\para{DNS}
Only two countries (out of the 186 tested) appear to censor DNS bidirectionally:
China and Iran. Both of these countries appear to censor in IPv4 and IPv6 fully,
both in terms of the IP addresses we send to, and that they are able to
effectively block A and AAAA records.

\para{HTTP}
HTTP appears widely censored, though we note it requires faking the SYN and ACK of
a TCP handshake before sending the censored request, as these censors appear to
be stateful in their censoring of HTTP (e.g. they do stateful connection
tracking). Interestingly, many of these countries only censor HTTP in IPv4,
including Morocco, Tanzania, Kuwait, Lebanon, and Turkey. We note that for many
of these countries, IPv6 adoption is low, potentially explaining why these
censors have opted to only support IPv4 blocking. An exception to this is
Kuwait, which according to Akamai and Google has a 16-18\% IPv6 adoption
rate~\cite{akamai-ipv6,Google-IPv6}, suggesting that simply using IPv6 may be an
easy way to avoid censorship there.

\para{TLS}
Similar to HTTP, TLS is also widely censored bidirectionally, though
predominately statefully (i.e. we must send a TCP handshake to be censored).
Only China censors TLS without a TCP handshake. With the exception of Lebanon,
all of the countries that censor TLS support IPv6 as well as IPv4.

\subsection{Case Studies}
\label{sec:prevalence:case}

\para{China}
censors bidirectionally for all 3 protocols that we tested (DNS, HTTP, and TLS).
We queried a total of 104k IPv6 and 91k IPv4 addresses in China. For stateful
HTTP and stateful TLS, we received a censored response from  $>73\%$ of IPv4
addresses and $66>\%$ of IPv6 addresses. However, for the stateless counterparts
of these protocols, we received a lot less censored responses. For HTTP and TLS
stateless protocols, we received censored responses from only 22\% of IPv4
addresses and 8\% from IPv6 addresses. For all the tested protocols, except
stateless HTTP, a majority ($>95\%$) of the censored responses we received were
\textit{RST} or \textit{RST + ACK} packets (for connection tear down), a
signature of China's Great
Firewall~\cite{bock2019geneva,wang2017your,weaponizing-middleboxes}
However for stateless HTTP, $>80\%$ of the censored responses we received were HTTP injections.

We found that for DNS probes carrying either A or AAAA queries China censors at
a similar rate, $87.2\%$ of the tested IPv4 addresses and $99.3\%$ of IPv6
addresses. We note that there were no responses to our DNS control probes for
IPv6 in China.

During our preliminary scans, one of our IPv4 addresses used in
scanning was blocked by most of China, resulting in a large drop in our results. We
rectified this by scanning from a different vantage point for just China's IPs
(and confirming this new vantage point remained unblocked), and we report on
those numbers in this paper. However, we have observed this type of blocking on
two of our IPv4 addresses that have scanned for censorship in China, and we speculate that
this may be a
feature of China's Great Firewall aimed at preventing the study of it. But this
feature appears limited to IPv4: our IPv6 address remained unaffected by these policies,
despite sending the same experiments to an even greater number of IPs inside
China.

\para{Iran} censors nearly all of the protocols we studied, and supports IPv4 and IPv6
for most of them. Interestingly, it appears that Iran only blocks TLS bidirectionally
over IPv6; however during our study Iran made several changes to their routing in response to protests, making it difficult to know if this is related.
Despite only have a 3-8\% IPv6
adoption rate~\cite{akamai-ipv6,Google-IPv6}, Iran appears to have a
fully-functional IPv6
censorship system. Iran also censors stateful
HTTP bidirectionally. We received a censored response for 50\% of IPv6 and 64\%
of IPv4 addresses. All of the censorship responses from Iran were \textit{RST}
packets \textbf{and} HTTP block pages. 

Among stateful and stateless TLS, we only found evidence of bidirectional censorship of stateful TLS over IPv6. 33\% of all IPv6 addresses sent back a censorship response to our stateful TLS queries. However, preliminary experimentation showed that Iran censors stateless and stateful HTTP and TLS at similar rates. %

\para{Russia}
While it is known that Russia censors its
Internet~\cite{ramesh2020decentralized}, we found only negligible evidence of
bidirectional censorship at the national level for Russia. Among all protocols,
stateful HTTP had the highest rate of bidirectional censorship at only 7\% of
the 121k IPv4 and only 3\% of 7.7k IPv6 addresses receiving a censored response.

We followed up to determine why we were not able to see widespread censorship,
and found this is largely due to the way that Russia censors at individual ISPs,
rather than at the edge of their network or IXPs, making it less likely to
function bidirectionally. We analyze our scans against
Quack~\cite{vandersloot2018quack}, a technique that does
detect unidirectional censorship employed in Russia in
Section~\ref{sec:compare}.

\para{Tanzania} censors bidirectionally for HTTP (stateful and stateless) but
for none of the other protocols we tested. We received a censored response for
$>35\%$ of all IPv4 addresses we queried in Tanzania. Almost all the censored
responses we received were HTTP block-pages. Upon following up on some of the
censored IP addresses by sending HTTP requests for domains that were censored
for Tanzania, we found reference to the \textit{Tanzania Cyber Crimes
Bill}~\cite{Tanzania45:online}, confirming the censorship was part of the
national firewall. Tanzania has only minimal IPv6 adoption
(0.3-0.4\%~\cite{akamai-ipv6,Google-IPv6}), and none of the 40 IPv6 addresses we
tested showed evidence of censorship.

\subsection{Missing countries}
\label{sec:compare}

\TabQuack

We note that while our bidirectional censorship measurement technique is able to
be easily applied globally to measure both IPv4 and IPv6 censorship, there are
notably several countries that we do not observe censorship in, but that prior
work has identified as censoring. For instance,
Russia~\cite{ramesh2020decentralized} and India~\cite{singh2020india} both
censor the protocols we looked for, yet our results suggest very minimal
censorship there.

We hypothesized this is due these countries censoring in a way that is not
bidirectional. For instance, in Russia, since censorship is often done at
individual ISPs close to the end user, the censorship can be done in a way that
only impacts the users at that ISP. Prior work has shown that much of Russia's
censorship only applies to connections that originate from within the country,
making it invisible to our technique~\cite{xue2021throttling}.

To investigate this hypothesis, we compare our results to that from Quack, a
technique that leverages echo servers in censoring countries to test HTTP (and
TLS) censorship~\cite{vandersloot2018quack}. Since Quack's censorship-triggering traffic will pass in both
directions past a censor, it is capable of measuring unidirectional censorship.
However, since it relies on finding echo servers, it is limited to IPv4
addresses, since it is infeasible to scan for similar echo servers in IPv6.

Comparing our IPv4 results to that from Quack in Table~\ref{tab:quack}, we see that for many countries,
we see significantly lower censorship rates, even in the same subnets that host
Quack echo servers that see censorship. This implies that much of the missing
countries (e.g. Russia, India) may be due to the way these countries censor, and
a limitation of our technique.

\section{Related Work}\label{sec:related}

Prior targeted censorship measurement studies contribute to a better understanding of
block-list infrastructure~\cite{ramesh2020decentralized, USESEC21:GFWatch} and
have helped to explain blocking phenomena~\cite{global2002great, Anonymous2020:TripletCensors}.
Meanwhile, global studies have also yielded higher-level views on the use of DNS censorship
around the world
\cite{vandersloot2018quack, scott2016satellite,
pearce2017global, sundara2020censored, niaki2020iclab}. However, all of these
studies have required Internet-wide scans, that are only feasible on IPv4. Thus,
there is a gap of knowledge when it comes to IPv6 censorship. Our work performs a
global measurement of DNS, HTTP, and TLS censorship through the lens of
comparing the
differences in IPv4 and IPv6 censorship deployments around the world.

While prior global censorship measurement work has been limited to IPv4, 
there have been several efforts to incorporate IPv6 or understand how censors
deal with IPv6-specific features.
In March 2020 Hoang \etal collected DNS records injected by the Great
Firewall in order to classify the addresses provided, block-pages injected, and
the set of hostnames that receive injections~\cite{USESEC21:GFWatch}. Their
analysis investigates the commonality of addresses injected by the GFW, finding
that all injected \texttt{AAAA} responses are drawn from the reserved teredo
subnet \texttt{2001::/32}. However, because this study does not directly
compare the injection rates of A vs AAAA or differences in injection to DNS
queries sent over IPv4 versus IPv6, our efforts complement their findings and
provide a more detailed understanding of IPv6 censorship in China.
A 2021 investigation of HTTP keyword block-lists associated with the Great
Firewall found that results are largely the same between IPv4 and
IPv6~\cite{weinberg2021chinese} using a single vantage point in China that had
both IPv4 and IPv6 connectivity. Their results corroborate ours---that China does
censor over IPv6. However, the authors note that over IPv6
connections, the the Firewall failed to apply its signature temporary 90~second
``penalty box'' blocking subsequent connections between the two hosts described
by numerous previous studies~\cite{xu2011internet,clayton2006ignoring}. This
supports our finding that at least some parts of the GFW's infrastructure supporting IPv4 and
IPv6 are implemented and/or deployed independently.

\section{Discussion and Conclusions} \label{sec:discussion}

In this section, we detail limitations
(\Cref{sec:discussion:limitations}), directions for future research
(\Cref{sec:discussion:future}), and the takeaways of our work.
(\Cref{sec:discussion:conclusions}).

\subsection{Limitations}
\label{sec:discussion:limitations}
Measuring censorship in order to gain an understanding of the underlying
infrastructure and identify weaknesses for circumvention is a challenging task
due to the absence of ground truth for validation and the often probabilistic
nature of censorship and networking failures which are easily confused.

Although we take care to always err on the side of caution and consider many
confounding factors including end-point type and AS diversity, our work is
fundamentally a best-effort attempt at trying to identify the gaps that have
emerged in Bidirectional censorship deployments because of the increased
adoption of IPv6.

\para{External sources of data.}
Our study relies on multiple data sources including \textit{The Citizen
Lab}~\cite{TheCitiz6:online} for our domain lists, the \textit{Route Views
Project}~\cite{RouteVie20:online} for BGP allocation data, and Maxmind's
datasets~\cite{maxmind-connectiondb} for geolocating our chosen target
addresses. Although each of these datasets has been validated in the past and
are commonly used in research, our results and their corresponding analyses are
limited by their reliability.

\para{Network stability}
Due to the nature of bidirectional censorship, where the typical benign response
is no response, it is not possible to easily distinguish a negative result from
a probe that would receive a censorship response but was dropped by the network
before it reached the censoring link. We believe that our results are still
representative as we are looking for the presence of bidirectional censorship
capabilities in aggregate rather than relying on the correctness each individual
probe. The drop rates in most networks are low and each allocation has 10
selected addresses, each of which receives a probe for each of the 1400 domains
that we test providing a significant level of redundancy.

\subsection{Future Work} \label{sec:discussion:future}

While we find that a global scan of bidirectional censorship provides a broad
view of network interference, several open questions and opportunities for
further investigation still exist.

\para{Other Protocols}
We present results for several protocols well known to be censored at large
scales around the world, however this study is in no way an enumeration of
censorship strategies. For example, HTTP keyword based censorship is a common
strategy known to be deployed in several nation-state networks. However, the
keyword blocklists tend to be more regionally specific and significantly larger.
At the scale of target addresses that we send in this work the number of probes
becomes difficult to manage.

Along with changes relating IP versions the protocols that carry commodity
traffic change over time, as protocols are updated and improved. To this end we
did a global measurement of both Quic and DTLS (both of which are UDP variants
of the TLS protocol) by placing the domain under test in an SNI extension of a
ClientHello packet equivalent to our TLS probe. Neither protocol showed strong
signs of censorship relating to the server name.

Our Quic probes received responses from 8233 ($0.27\%$) and 1318 ($0.33\%$)
addresses respectively for IPv4 and IPv6 respectively. All of the responding
IPv6 addresses belong to cloud hosting providers Cloudflare, Fastly, and
NextDNS. A large number of the remaining IPv4 addresses were geolocated to US,
which is consistent with both the scale of allocations in the US and the
location of the parties associated with the development of the Quic protocol.
For DTLS 1396 addresses ($0.04\%$) responded to any probe all of which were in
IPv4. Most of the responding addresses were in AS 2044 which is associated with
a company offering hosting/connectivity as a service, and AS 4193 which is
associated with the State of Washington in the US. For both Quic and DTLS the
number of addresses that responded to experimental probes, but not control
probes was so small that it could be attributed to network instability or other
statistical error. We interpret this as indicative of no current bidirectional
censorship of either protocol relating to the SNI extension.

\para{Circumvention opportunities}
The incongruity that we find in censorship deployments demonstrates that there
may be opportunities to leverage the gap to circumvent network based limitations
on free speech. Censorship efficiency and distribution is not one-to-one between
IPv4 and IPv6 allowing for potential chosen path attacks for example.
Furthermore, while not explored in this work it may be possible that protocols
designed for interim or transition period between IPv4 and IPv6, such as 6-to-4
tunnelling and teredo, would go unseen by censors.

\para{Fingerprinting}
Given the large number of networks and network-actors that we measure in this
work we intend to perform a classification of censorship behaviors at the
protocol level to identify and link common censorship infrastructure and
implementation commonalities where circumvention techniques can be shared
laterally.

For target addresses that are identified as having a censor on-path, follow-up
scans using tools such as geneva~\cite{bock2019geneva} could be done to further
explore the extent to which censorship can be fingerprinted and circumvented.
Elements of such a fingerprint would include packet level details like IPID and
IPTTL of injected packets as well as censorship trigger conditions relating to
protocol validity elements like flags, checksums, and extensions.

\para{Intentional Packet Drops}
One key censorship response that we do not capture in the work is intentional
packet drops. This is not passively differentiable from the benign response in
our scan, however it is a widely deployed censorship technique. One potential
way to bridge this gap is to extend this work to measure packet drops by
incorporating an analysis of the IPID in response packets sent by the truly
benign target addresses. For targets that send TCP RST packets with a globally
incrementing IPID shared by all destination hosts analysis can indicate when a
packet to the target address was dropped in-flight as described by Ensafi
\etal~\cite{ensafi:detecting}. Again this type of analysis increases the number
of packets required to establish confidence due to noise and network
instability, but such a measurement would provide a significant extension to the
results we present in this work.

\subsection{Conclusions} \label{sec:discussion:conclusions}

Many governments continue to censor the Internet.
In order to better understand the scope of this censorship, particularly with respect
to the ongoing deployment of IPv6 is effected,
we perform a global measurement of bidirectional censorship on both the IPv4 and IPv6
Internet.

We experimentally find that many networks deploy at least one form of
bidirectional censorship capability. We spotlight several countries that censor at
a seemingly national scale, and capture measurements implicating several more.
In particular, we find that while some censors support IPv6, there are others that only
censor in IPv4. In addition, there are differences between the fraction
of networks that censors can employ blocking in the respective IP versions: some
censors block more networks in IPv4, for instance, suggesting that some users
may be able to escape some or all forms of censorship simply by using IPv6 if available.

It is important to understand the current state of censorship in the context of
a developing Internet. This work contributes to a broader understanding of
global censorship and the gaps therein.

\bibliographystyle{plain}
\bibliography{censorship}

\end{document}